\documentclass[10pt,conference]{IEEEtran}

\usepackage[utf8]{inputenc}
\usepackage{cite}
\usepackage{amsmath,amssymb,amsfonts, amsthm}
\usepackage{dsfont}
\usepackage{mathtools}
\usepackage{units}
\usepackage[flushleft]{threeparttable}
\usepackage{enumitem}
\usepackage{booktabs}
\usepackage[noend]{algorithmic}
\usepackage[ruled,linesnumbered, noend]{algorithm2e}
\usepackage{scrextend}
\usepackage{booktabs}
\usepackage[table,xcdraw]{xcolor}

\usepackage[subrefformat=parens,labelformat=parens,caption=false,font=footnotesize]{subfig}

\usepackage{tikz}
\usetikzlibrary{shapes,arrows}
\usetikzlibrary{calc}

\def\Gbps{~\mathrm{Gbps}}

\def\GHz{~\mathrm{GHz}}
\def\MHz{~\mathrm{MHz}}
\def\Hz{~\mathrm{Hz}}

\def\m{~\mathrm{m}}

\def\dB{~\mathrm{dB}}
\def\dBi{~\mathrm{dBi}}
\def\dBm{~\mathrm{dBm}}

\newcommand{\mathrmbold}[1]{\boldsymbol{\mathrm{#1}}}

\def\tablescale{0.9}

\newcommand{\wrt}{\textit{w.r.t.~}}

\newcommand{\ie}{\textit{i.e.~}}

\newcommand{\eg}{\textit{e.g.~}}

\newcommand{\titleheader}{This work has been accepted for publication in 2023 IEEE IEEE International Symposium on Personal, Indoor and Mobile Radio Communications (PIMRC): Cellular Networks Track}

\makeatletter
\def\ps@headings{%
\def\@oddhead{\mbox{}\scriptsize \titleheader}%
\def\@oddfoot{\scriptsize \@date \hfil }%
}

\def\ps@IEEEtitlepagestyle{%
\def\@oddhead{\mbox{}\scriptsize \titleheader \rightmark \hfil }%
}
\makeatother

\title{DEDICAT-EUCNC23-Hierarchical Multi-MAP cooperation for reconfigurable 5G dynamic networks}
\title{DEDICAT-EUCNC23-Hierarchical Deep Reinforcement Learning for reconfigurable 5G dynamic networks via Multi-UAV cooperation}
\title{Learning Reconfigurable Cooperative Multi-UAV 5G networks via Federated Multi-Agent Deep Reinforcement Learning}

\title{Federated Multi-Agent Deep Reinforcement Learning for Dynamic and Flexible 3D Operation of 5G Multi-MAP Networks}

\author{\IEEEauthorblockN{Esteban Catté, Mohamed Sana and Mickael Maman}

\IEEEauthorblockA{ CEA-Leti, Universite Grenoble Alpes, F-38000 Grenoble, France \\
\{esteban.catte, mohamed.sana, mickael.maman\}@cea.fr\\}
}

\date{November 2022}

\begin{document}

\maketitle

%==============================================================================================%
%                                         ABSTRACT                                          %
%==============================================================================================%

\begin{abstract}

This paper addresses the efficient management of Mobile Access Points (MAPs), which are Unmanned Aerial Vehicles (UAV), in 5G networks. We propose a two-level hierarchical architecture, which dynamically reconfigures the network while considering Integrated Access-Backhaul (IAB) constraints. The high-layer decision process determines the number of MAPs through consensus, and we develop a joint optimization process to account for co-dependence in network self-management. In the low-layer, MAPs manage their placement using a double-attention based Deep Reinforcement Learning (DRL) model that encourages cooperation without retraining. To improve generalization and reduce complexity, we propose a federated mechanism for training and sharing one placement model for every MAP in the low-layer. Additionally, we jointly optimize the placement and backhaul connectivity of MAPs using a multi-objective reward function, considering the impact of varying MAP placement on wireless backhaul connectivity.
\end{abstract}

\begin{IEEEkeywords}
Mobile Access Points, Integrated access backhaul, Multi-agent Deep Reinforcement Learning, Federated Learning, mmWave Communications, Dynamic 5G Networks.
\end{IEEEkeywords}

%==============================================================================================%
%                                         INTRODUCTION                                          %
%==============================================================================================%

\section{Introduction}
5G aims to offer fair opportunities for User Equipments (UE) regardless of their location or mobility via efficient management. Mobile Access Points (MAPs), which are Unmanned Aerial Vehicles (UAV), are gaining attention as a flexible infrastructure, useful for various applications \cite{Maman2022}. MAPs can collaborate to form a Multi-MAP network, but there is limited research on managing them in dynamic networks with user mobility, interference, varying traffic, and fluctuating MAP numbers.
%==============================================================================================%
%                                         SOTA                                          %
%==============================================================================================%
Our objective is to efficiently manage multiple MAPs in terms of their number, placement, and trajectory while considering dynamic constraints over a longer time scale than the current state-of-the-art approaches. Previous studies have explored different approaches leveraging the 3-dimensional (3D) mobility of MAPs, but often without accounting for all the dynamic network constraints simultaneously. For instance, in \cite{Peer2021}, the authors proposed an iterative optimization method for MAP placement based on user mobility. Another study by Ghanavi et al. \cite{Ghanavi2018} extended the scenario to multiple MAPs managed by a reinforcement Q-learning algorithm. Wang et al. \cite{Wang2022} introduced a virtual forces algorithm based on statistical user distributions for computing network cartography. It is worth noting that user distribution can impact MAP numbers and deployment positions, even when the number of UEs remains constant. These diverse solutions demonstrate the variety of MAP management techniques, highlighting the need for iterative approaches to efficiently handle dynamic network constraints. However, ensuring long-term performance in a constantly changing network remains a challenge.

The aforementioned papers highlight the potential of using a greedy MAPs deployment approach to determine their optimal number. For instance, in \cite{Sharafeddine2018}, \cite{Sabzehali2021}, \cite{Zhang2021}, \cite{Lyu2017}, \cite{Qin2019}, proposed solutions adjust the number of deployed MAPs iteratively to meet network constraints. However, this approach may suffer from convergence delays and does not account for network evolution. In contrast, our study proposes a hierarchical architecture that dynamically determines the number of MAPs for user coverage, independent of the placement procedure. Our architecture aims to strike a balance between cost and coverage by determining both the number and positions of MAPs, as these aspects affects each other.

Obviously, MAP management must adapt to changing network conditions, including trajectory adjustments. In \cite{Wu2018}, authors used a successive convex optimization to optimize MAP trajectories and UE data rates under mobility constraints. However, a significant breakthrough in MAP trajectory optimization has been achieved with Multi-Agent Deep Reinforcement Learning (MADRL) models. In \cite{Zhao2020} and \cite{Qin2021}, authors proposed target MADRL models based on the actor-critic architecture to handle multiple factors. Authors of \cite{Zhou2022bis} proposed a MADRL approach with pre-deployed MAPs on UE clusters. This approach takes advantage of the low-complexity deployment algorithm and the ability of MADRL model to adjust positions in complex environments. 

Our paper presents a problem formulation and proposes a two-level hierarchical architecture based on joint optimization for a dynamic 5G network while considering Integrated Access-Backhaul (IAB) constraints. The decision process is scalable and distributed and it determines the number of MAPs through consensus in the high-layer. In the low-layer, MAPs manage their placement using our previously proposed dual-attention based DRL model \cite{Catte2023} that encourages cooperation without any a-priory information or retraining procedure. To increase the generalization ability of learned model, reduce complexity and improve performance in novel scenarios, we propose a federated mechanism that involves training and sharing one placement model for every MAP, as suggested in \cite{Hu2020}.

Additionally, we aim to jointly optimize backhaul connectivity of MAPs using a multi-objective reward function, considering the impact of varying MAP placement on wireless backhaul link as highlighted in previous studies \cite{Iradukunda2021} and \cite {Dai2021}.

The paper is organized as follows. Section II presents the system model and Section III formulates the addressed problem. Then, Section IV describes our proposed solution, whereas Section V provides our numerical results. Finally, Section VI concludes the paper.

%==============================================================================================%
%                                         SYSTEM MODEL                                         %
%==============================================================================================%

\section{System Model}
\begin{figure}[!t]
\includegraphics[width=\columnwidth]{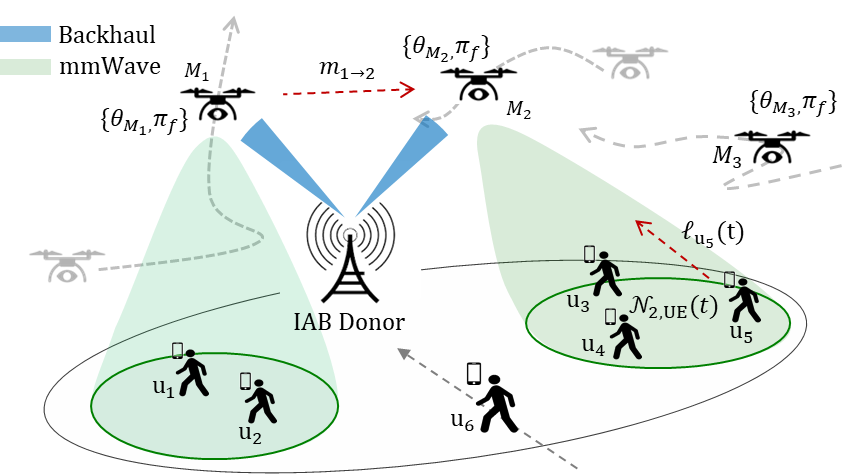}
\caption{System model with one IAB Donor, two deployed MAPs maintaining their trade-off value $\{\theta_{M_1}, \theta_{M_2}\}$ and sharing policy $\pi_f$ with one joining MAP, five communicating UEs with one joining UE and corresponding links.}
\vspace{-0.1cm}
\end{figure}

We consider a downlink network composed of $M$ MAPs operating at mmWave frequencies. Each flying MAP can establish a backhaul link with a grounded IAB donor. We define $M_s(t)$ as the number of deployed MAPs at time $t$, which move to provide services to $K(t)$ UEs. Let $\mathcal{U}(t) = \{1,\dots, K(t)\}$ be the set of UEs, $\mathcal{S}_0(t)$ the set of all Base-Station (BS) including the IAB donor indexed by 0 and $\mathcal{S}(t) = \{1, \dots, M\}$ denotes the dynamic set of deployed MAPs. %\footnote{For sake of clarity, we adopt $\mathcal{S}$ and $\mathcal{S}_0$ notation for the rest of the paper}. 
We assume that UEs can be associated with only one MAP $i\in\mathcal{S}_0(t)$ providing the maximum signal-to-noise ratio (via \texttt{max-SNR} algorithm). In our system model, we assume that the grounded location $\ell_j(t)\in\mathbb{R}^2$ of UEs changes with time, requiring dynamic and on-demand reconfiguration of MAPs deployment. Once deployed, MAP $i\in\mathcal{S}(t)$ can adapt its 3D location $\ell_i(t)$ in a region $\mathcal{L}$ of $\mathbb{R}^3$ space and can only serve at most $K_i(t)$ UEs due to limited beamforming capability. In this dynamic network, optimizing the number and placement of MAPs is a challenging and important task to improve network spectral efficiency. Indeed, MAPs should dynamically adjust their number and location to follow UE's dynamics while limiting interference.

%==============================================================================================%
%                                         CHANNEL MODEL                                        %
%==============================================================================================%
\subsection{Channel Modeling}
Our system model considers an \textit{out-of-band} relaying IAB network where the access and backhaul links are orthogonal and do not interfere on each other. In this context, we split the available mmWave bandwidth $B$ into parts dedicated to backhaul ($\mu B$) and access network ($(1-\mu) B$), where $\mu\in[0,1]$. We assume that both the access and backhaul links use Spatial Division Multiple Access (SDMA). Thus, when UE $j$ is receiving data from BS $i$, it
%With all the bandwidth allocated, UE $j$ receives data from BS $i$ at rate $R_{i,j}$ and 
experiences a downlink signal-to-interference-plus-noise ratio $\mathrm{SINR}^{(a)}_{i,j}$, which reads as:
\begin{align}\label{eq:interf_acess}
    \mathrm{SINR}_{i,j}^{(a)}(t) = \frac{\zeta_{i,j}(t) P_{i,j}^{\rm Tx} G_{i,j}^{\rm Tx}(t) G_{i,j}^{\rm H}(t) G_{i,j}^{\rm Rx}(t) }{I_{i,j}^{(a)}(t) + (1-\mu)N_0B}.
\end{align}
Here, $P_{i,j}^{\mathrm{Tx}}$ is the transmit power from BS $i$ towards UE $j$, $N_0$ is the Gaussian noise power spectrum density. Also $G_{i,j}^{\mathrm{Tx}}(t)$ and $G_{i,j}^{\mathrm{Rx}}(t)$ are the transmit and receive antenna gain between BS $i$ and UE $j$, respectively. To reflect the impact of the environment on channels, we define $\zeta_{i,j}(t)$ as the small-scale fading coefficient, $G_{i,j}^{\mathrm{H}}(t)$ channel gain capturing the path-loss and large-scale shadowing effect. Eventually, $I_{i,j}^{(a)}(t)$ is the total intra- and inter-cell interference experienced by UE $j$ communicating with BS $i$. Hence, the access capacity, $C_{i,j}^{(a)}(t)$, of the link between BS $i$ and UE $j$ reads as:
\begin{equation}\label{eq:capa_acess}
    C_{i,j}^{(a)}(t) = (1-\mu)B\cdot \mathrm{log}_2(1+x_{i,j}(t)\mathrm{SINR}_{i,j}^{(a)}(t)),
\end{equation}
where $x_{i,j}$ is the binary UE association variable, which equals $1$ when UE $j$ is associated with BS $i$ and $0$ otherwise.
Similarly, the backhaul capacity, $C_{i}^{(b)}(t)$, of the link between MAP $i$ and the IAB donor reads as:
\begin{equation}\label{eq:capa_backhaul}
    C_{i}^{(b)}(t) = \mu B \cdot \mathrm{log}_2(1+z_{i}(t)\mathrm{SINR}_{i,j}^{(b)}(t)),
\end{equation}
where we define $z_i(t)$ as the binary backhaul link association variable, which indicates if a MAP is currently deployed or not.  Here, the $\mathrm{SINR}_{i}^{(b)}(t)$ experienced by the MAP $i$ communicating with the IAB donor is given by:
\begin{equation}\label{eq:interf_backhaul}
    \mathrm{SINR}_{i}^{(b)}(t) = \frac{\zeta_{0,i}(t) P_{0,i}^{\rm Tx} G_{0,i}^{\rm Tx}(t) G_{0,i}^{\rm H}(t) G_{0,i}^{\rm Rx}(t) }{I_{i}^{(b)}(t) + \mu N_0B},
\end{equation}
where, $I_{i}^{(b)}(t)$ denotes the intra-backhaul interference. %and self-interference coming from transmitted and received messages with MAPs. 

It is worth noting that in Eq. \eqref{eq:capa_acess}-\eqref{eq:capa_backhaul}, the $\mathrm{SINR}$ and the channel capacity depend on path losses and interference influenced by various topological factors. Our system model considers ground-to-ground %sub-6GHz path loss 
and air-to-ground mmWave path loss, which are affected by Line-of-Sight (LoS) conditions and the distance $d_{i,j}(t)=\|\ell_i(t) - \ell_j(t)\|$  between MAP $i$ and UE $j$ at time $t$. We omit full description here due to lack of space and refer readers to our previous work \cite{Catte2023}.

\subsection{Effective Rate and Network Sum-rate}
Let $D_j(t)$ define the traffic request of UE $j$ at time $t$ (in $\mathrm{bps}$). Hence, $\Gamma_{i,j}(t)= \mathrm{min}(D_j(t), C_{i,j}^{(a)}(t))$ represents the effective data requirement on the access link between UE $j$ and MAP $i$. Thus, if $\beta_{i,j}(t) \in [0,1]$ is the fraction of MAP $i$ backhaul capacity $C_i^{(b)}(t)$ allocated to UE $j$, %For each served UE $j$, we define $\Gamma_{i,j}(t)= \mathrm{min}(D_j(t), C_{i,j}^{(a)}(t))$ the effective data requirement when associated to BS $i$ with $D_j(t)$ the UE $j$ traffic request at time $t$ (in $\mathrm{bps}$).
the instantaneous effective rate $R_{i,j}(t)$ perceived by UE $j$ from BS $i$ reads as:
\begin{align}\label{eq:eff-rate}
        R_{i,j}(t) = \begin{dcases}	 \mathrm{min}(\Gamma_{i,j}(t), \beta_{i,j}(t) z_i(t) C_i^{(b)}(t)),\forall i \in \mathcal{S},\\
				\Gamma_{i,j}(t),\hspace{3.85cm}\text{if $i=0$}.
				    \end{dcases}
\end{align}
Finally, we define the total network sum-rate $R(t)$ as:
\begin{equation}
    R(t) = \sum_{i\in\mathcal{S}}\sum_{j\in\mathcal{U}(t)} R_{i,j}(t)
\end{equation}

%==============================================================================================%
%                                         PROBLEM                                              %
%==============================================================================================%
\section{Problem Formulation}
Our goal is to optimize the user experience in this dynamic networks with varying traffic demand, locations, and numbers of MAPs and UEs. We aim to optimize at the same time i) the number of deployed MAPs, ii) their backhaul allocation and iii) the dynamic placement of each MAP. To do so, we formulate the Multi-MAP management problem to maximize the long-term sum rate as follows:
\begin{align}
	\underset{\mathrmbold{\Psi}(t)}{\mathrm{max}}~~& \lim_{T\rightarrow+\infty}\frac{1}{T}\sum_{t=1}^T \mathbb{E}[R(t)], \tag{$\mathcal{P}$} \label{eq:P}\\[-0.15cm]
    	\mathrm{s.t.~}~ & x_{i,j}(t), z_i(t) \in \{0,1\}, & \forall i\in \mathcal{S}_0, j\in \mathcal{U}(t), \tag{$\mathcal{C}_1$} \label{eq:C1}\\
		{}&\sum_{j \in \mathcal{U}(t)}x_{i,j}(t) \leq K_{i}(t), & \forall i \in \mathcal{S}_0(t), \tag{$\mathcal{C}_2$} \label{eq:C2}\\
		{}&\sum_{i \in \mathcal{S}_0}x_{i,j}(t) \leq 1, & \forall j \in \mathcal{U}(t), \tag{$\mathcal{C}_3$} \label{eq:C3}\\
		{}&M_s(t)=\sum_{\mathclap{i \in \mathcal{S}}}z_i(t)\leq M, & \tag{$\mathcal{C}_4$} \label{eq:C4}\\
    	&\beta_{i,j}(t) \in [0, 1], &  \forall i\in \mathcal{S},j\in \mathcal{U}(t), \tag{$\mathcal{C}_5$} \label{eq:C5}\\
    	{}&\sum_{j \in \mathcal{U}(t)}\beta_{i,j}(t) \leq 1, & \forall i \in \mathcal{S}(t), \tag{$\mathcal{C}_6$} \label{eq:C6}\\
    	{}&\ell_i(t)\in\mathcal{L}\subset\mathbb{R}^3, & \forall i \in \mathcal{S}(t), \tag{$\mathcal{C}_7$} \label{eq:C7}\\
    	{}&||\ell_i(t+1)-\ell_i(t)|| \leq \Delta \ell, & \forall i \in \mathcal{S}(t), \tag{$\mathcal{C}_{8}$} \label{eq:C8}%\\
    	%{}&0 \leq M_{s}(t) \leq M, \tag{$\mathcal{C}_{9}$} \label{eq:C9}
\end{align}

where $\mathrmbold{\Psi}(t) = \{z_i(t), \beta_{i,j}(t), \ell_i(t), \forall i,j\}$ are the optimization variables and the expectation in \eqref{eq:P} is taken \wrt the random processes, whose statistics are unknown. In \eqref{eq:P}, constraint \eqref{eq:C2} ensures that each BS $i$ serves at most $K_i(t)$ UEs simultaneously. Constraint \eqref{eq:C3} guarantees that each UE is associated to one BS at a time. Similarly, \eqref{eq:C4} ensures that the IAB donor serves at most $M$ active backhaul links simultaneously. Moreover, \eqref{eq:C5}-\eqref{eq:C6} guarantees a positive backhaul allocation $\beta_{i,j}(t)$ for each UE $i$ connected to MAP $i\in\mathcal{S}(t)$ and sum to at most one at each time $t$. Finally, regarding MAPs mobility, \eqref{eq:C7}-\eqref{eq:C8} define a bounded region $\mathcal{L}$ of space where MAPs cannot move more than $\Delta \ell$ meters at a time. %Finally, constraint \eqref{eq:C9} ensures to not activate more than MAPs $M$ at the same time. 
Problem \eqref{eq:P} is a non-convex combinatorial problem whose complexity increases with network size. In addition, there is an interdependence in optimization variables. Indeed, the required number of MAPs depends on UE topology, such as location and traffic demand distribution, which determines whether a dense or scattered deployment is necessary. For determining the optimal MAP locations, a centralized exhaustive search is not feasible due to interdependence between the number and locations of MAPs and the complexity of the network's interference profile. Following \cite{sana2023}, given the user association $x_{i,j}(t)$, here specified by the \texttt{max-SNR} algorithm, the backhaul capacity allocation $\beta_{i,j}(t)$ can be obtained using convex optimization. Using \texttt{max-SNR} algorithm, the values of $x_{i,j}(t)$ are defined by the locations and number of MAPs. Therefore, in the remaining, we focus on finding $\{z_i(t), \ell_i(t)\}$. To solve this problem with limited complexity, we propose a two-level hierarchical optimization framework to optimize $\mathrmbold{\Psi}(t)$.

%==============================================================================================%
%                                   PROPOSED SOLUTION                                          %
%==============================================================================================%
\section{Proposed Solution}

\begin{figure}[!t]
\includegraphics[width=\columnwidth]{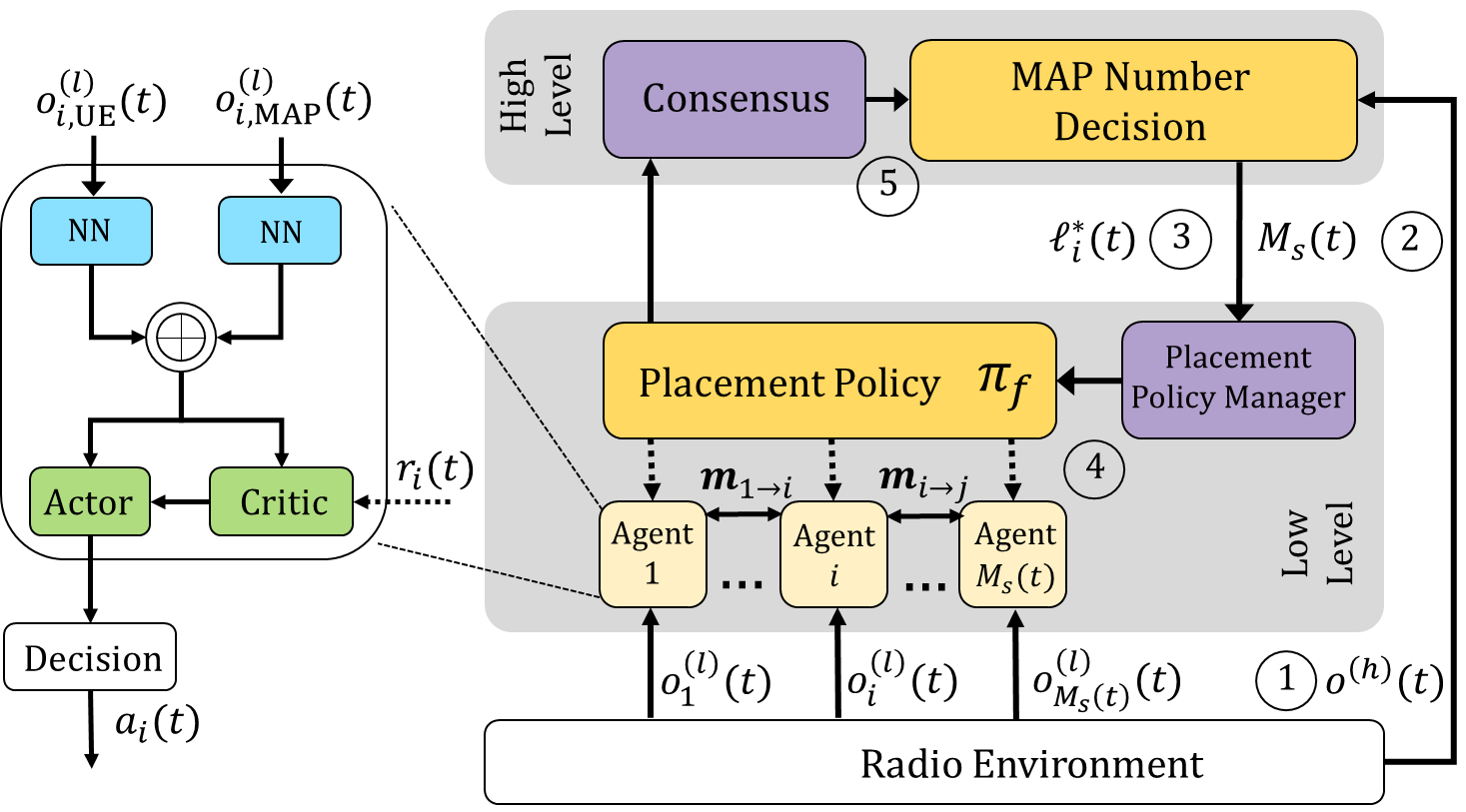}
\caption{Proposed hierarchical architecture for 3D operation of MAPs.}
\label{fig:framework}
\vspace{-0.2cm}
\end{figure}

We propose a two-level framework to address problem \eqref{eq:P}, wherein the high-level is responsible for jointly determining the number of MAPs and the low-level uses federated MADRL to position the MAPs under time-varying network constraints. 
As illustrated in Fig. \ref{fig:framework}, the high-level gathers network information $o^{(h)}(t)$ (1) and feedbacks of MAPs currently deployed to make the \textbf{MAP number decision} (2). During the training phase, the high-level determines the best target location $\ell_i^{*}(t)$ for each MAP $i$, which is used to compute MAPs agent training rewards (3). These locations are no longer transmitted afterwards, \ie  they only serve for training, since the agents have learned to determine them on their own. Then, the low-level \textbf{placement policy manager} loads the \textbf{placement policy} to the deployed MAPs and sends them to the network. For the training phase, each MAP federates its local model to a common placement policy $\pi_f$ (4). Finally, each MAP determines its relative importance within the network based on its trade-off to achieve a \textbf{Consensus}. Thus, each MAP decides whether to repatriate or enable a new MAP for assistance in serving the UEs (5). This dynamic deployment of MAPs is iterative, scalable, and distributed.

\subsection{High-Level - Decentralized Trade-off}
In this section, we discuss the intuitions and motivations behind the proposed distributed \textbf{\texttt{Trade-off algorithm}}. First, the MAP number issue must be addressed simultaneously with topology and radio configuration conflict constraints.
In fact, deciding the number of MAP to deploy depends on: i) the network \textbf{topology} \ie UE's distribution; ii) the network nodes \textbf{configuration} \ie UE's association; which are not considered in standard approaches like clustering. We propose to include both aspects in the calculation of the trade-off $\theta_i$ specific for each MAP $i$.
Thus, the associated problem raises multiple challenges: the self-dependency of MAP locations and number, ensuring sufficient UE coverage, maintaining backhaul connectivity, minimizing the number of MAPs to limit operational complexity.
To tackle these objectives, with low complexity and a long-term vision, we propose an iterative algorithm that considers both aspects in the trade-off calculation. 

\vspace{0.25cm}
\noindent
\textbf{Trade-off computation}. As described in Algorithm \ref{alg:toa}, each MAP $i$ maintains locally a trade-off value $\theta_i(t)$ to decide if it needs a support from a new MAP or to be repatriated. Therefore, decision making is decentralized and the number of MAPs is determined by consensus. Then, each MAP computes its UEs inertia $\varPhi_i(t)$, determined by the sum of squared distance of active UEs to the MAP, which captures the network topology surrounding the MAP. Additionally, the MAP $i$ determines if its beams are currently overloaded with served UE or lower than a threshold number of beam $K_{i,\min}$, which captures the current network configuration. We find here the interdependence of the optimization variables since the management of the beams is ensured by deciding $x_{i,j}$ and $z_i$.
When the inertia is high or when the station is overloaded, $\theta_i$ is increased for each aspect and decreased in the opposite cases. Both metrics are acquired via \textit{local monitoring} of UEs within its coverage range and averaged to guarantee a long-term vision. 
Notice that each MAP must set up and execute the low-level hierarchy related to the current network configuration, which implies an operational cost. In fact, the MADRL framework guarantees only a limited generalization ability as it must have constant size input information. In consequence, we propose a new approach based on a federated learning mechanism to unify every MAP model into a single model, easy to maintain.

\begin{algorithm}[!t]
    \SetAlgoLined
    \small
    \DontPrintSemicolon
    \SetNoFillComment
    \KwIn{
    $\mathcal{S}$ set of MAPs; $\mathcal{U}$ set of UEs\;
    $K_i$ maximum beam per MAP\;
    }
    Enable $K(t)/\mathbb{E}_i[K_i]$ MAPs and start low-level MADRL algorithm \cite{Catte2023}\;
    \For{$t \in [0, T_n]$}{
        \lIf{$t~\mathrm{modulo}~\tau_n=0$}{
            Init $\theta_i = \{i:0\} \forall i \in \mathcal{S}$
        }
        \For{$i \in \mathcal{S}$}{
            \lIf{$\overline{\theta_i} < 0$}{
                MAP $i$ decides to repatriate
            }
            \If{$\overline{\theta_i} > 0$}{
                MAP $i$ activates MAP $i^*$; $z_{i^*}=1$\;
                Placement Policy Manager sends $\pi_f$ to MAP $i^*$\;
            }
        }
        \If{$M_s(t) < M$}{
        Update backhaul capacity allocation $\beta_{i,j}$\;
        Update UE associations $x_{i,j}$\;
        \For{$i \in \mathcal{S}$}{
            Update $\theta_i$ through \textit{local monitoring}\;
            }
        }
    }
    \caption{\texttt{Trade-off Algorithm}}\label{alg:toa}
\end{algorithm}

%==============================================================================================%
%                                         Placement LEVEL                                   %
%=============================================================================================
\subsection{Low Level - Cooperative Placement}

\begin{table}[]
\vspace{-0.35cm}
\caption {Approaches Comparison}
\centering
\scalebox{\tablescale}{
\begin{tabular}{@{}|c|c|cc|@{}}
\toprule
\rowcolor[HTML]{C0C0C0} 
\textit{\textbf{Benchmark}} & \textbf{CODEBOOK} & \multicolumn{1}{c|}{\cellcolor[HTML]{C0C0C0}\textbf{Curriculum}} & \textbf{Federated} \\ \midrule
Context & Context-aware (Specialized) & \multicolumn{2}{c|}{Context-free (Generalized)} \\ \midrule
Complexity ($\mathcal{O}_c$) & $\frac{M(M+1)}{2}$ & \multicolumn{1}{c|}{$M$} & $1$ \\ \midrule
Policy & \begin{tabular}[c]{@{}c@{}}$\{\pi_{k,i}\}$,\\ $\forall k\in\{0,...,i\}, \forall i\in\mathcal{S}$\end{tabular} & \multicolumn{1}{c|}{$\{\pi_{i}\},\forall i \in \mathcal{S}$} & $\pi_{f}$ \\ \bottomrule
\end{tabular}}
\vspace{-0.3cm}
\label{tab:1}
\end{table}

To solve the dynamic MAP placement problem, we propose to model each MAP as an autonomous agent that have to cooperate to serve a dynamic 5G network. This approach comes with new challenges: follow and distribute UEs demand; schedule their path over time; collect and process surrounding information perception by their own. For this purpose, we propose a Multi-Agent Deep Reinforcement Learning (MADRL) algorithm as the low-level of our hierarchical architecture.
Thus, to efficiently solve the MADRL problem, we proposed in \cite{Catte2023} a double-attention actor-critic architecture. This model achieves a distributed cooperation without any prior information and without retraining procedures for time-varying scenarios. This cooperation is accomplished by learning, exchanging, and interpreting messages $\rm m_{i,j}$ between agents. The proposed solution solves multiple challenges: i) model-free property for the incoming radio environment; ii) agent state observations efficient representation; iii) network scalability; iv) distributed cooperation. 
In this approach, each MAP is modelled as an agent, which continuously learns to make autonomous decisions based on partial observations $o^{(l)}_{i,{\rm UE}}(t)$ from grounded UEs $\mathcal{N}_{i, {\rm UE}}$ and the messages $o^{(l)}_{i,{\rm MAP}}(t)$ received from other deployed MAPs $\mathcal{N}_{i, {\rm MAP}}$. We formalize this decision process as a Markov Decision Process (MDP). Each time slot, agents receive UEs location $\ell_j(t)$ and other MAPs location $\ell_i(t)$ to build a context representation $o^{(l)}_{i}(t)$. Based on this observation, agents select actions from a predefined set $\mathcal{A} = \{\text{forward, backward, up, down, left, right, hover}\}$, corresponding to movement of the associated MAPs along the selected direction with a fixed step size $\Delta \ell$. Agents then transition to new observations $o^{(l)}_{i}(t+1)$ and receive rewards according to the following multi-objective reward function:
\begin{align}\label{eq:reward}
    r_i(t) = (\delta_i(t)-1)d_{i}(t) + \delta_i(t) (C_{i}^{(b)}(t)-d_0).
\end{align}
Here, $\delta_i(t) = \mathds{1}(d_i(t) \leq d_0)$, where $d_0$ is a reference distance and $d_i(t) = \|\ell_i(t) - \ell_i^*(t)\|$ is the distance of MAP $i$ to its optimal location $\ell_i^*(t)$. Since this location is not known a priory, we approximate it during the training phase with the location of the nearest assigned centroid obtained by clustering UEs using \eg $\texttt{Kmeans}$ algorithm. As demonstrated in our previous work \cite{sana2023}, this multi-objective reward pushes agents to maximize user coverage and backhaul capacity at the same time. Each agent then learns a policy $\pi_i(t)$ that maximizes the expected sum of perceived ($\gamma$-discounted) rewards $\mathbb{E}_{\pi}[\sum_{\tau=t}^{T_l} \gamma^{\tau-t}r_{i}(\tau)]$ over a time horizon $T_l$, where $\gamma\in[0,1)$. However, as the dynamic of the network evolves, new training mechanisms are required to maintain network performance. Frequent training processes are prohibitive, induce latency, and a signalling overhead, which are detrimental to network operation efficiency. Therefore, new approaches are required to learn MAP placement policies, which are i) context-free \ie independent of the number of deployed MAPs $M_s(t)$, ii) scalable to cope with size-varying number $K(t)$ and position of UEs, iii) generalizable to different network deployment, which is a current and fundamental topic in MADRL, iv) and with limited operational complexity ($\mathcal{O}_c$).

\vspace{0.2cm}
\noindent
\textbf{Trivial approach via a codebook of policies.}
With the varying number of MAPs, the first trivial approach, which will serve as \textbf{\texttt{BASELINE}} is to maintain a representative set of scenario-specific policies to form a \textbf{\texttt{CODEBOOK}}. This approach devises specialized models for every combination of the number of deployed MAPs: $\{\pi_{k,i}\}$, $\forall k\in\{0,...,i\}, \forall i\in\mathcal{S}(t)$. Then, depending on the scenario, the \textbf{placement policy manager} selects the appropriate policy within the codebook to deploy. Obviously, this approach is complex and context-aware as it requires identification of the facing scenario, and maintaining $\mathcal{O}_c=\frac{M(M+1)}{2}$ different policies, where $M$ is the maximal number of MAPs. In addition, this approach may fail to generalize to unseen scenarios, which may not be captured by the codebook. Thus we propose to reduce the number of policies to maintain and at the same time increase their generalization ability. 

\vspace{0.2cm}
\noindent
\textbf{Share to conquer: a curriculum approach.}
In the context of a varying number of agents, we propose a curriculum MADRL (referred to \textbf{\texttt{C-MADRL}}) training approach. In contrast to the previous approach, each agent maintains its own model through all possible configurations, which reduces the operational complexity to $\mathcal{O}_c = M$. During the training procedure, we randomly sample a scenario with a random number of deployed MAPs. Then, the deployed MAPs cooperatively learn their respective policies, which we maintain across a different sampling of scenarios. This method is context-free and allows each MAP agent to generalize to different scenarios with a different number of deployed MAPs thus fostering cooperation with size-varying teammates.

\vspace{0.2cm}
\noindent
\textbf{A transferable policy via federated mechanism.}
Here, we propose a Federated MADRL (referred to \textbf{\texttt{F-MADRL}}) mechanism. The goal is to share the knowledge of placement and cooperation into a single policy $\pi_f$ that can be propagated to new any agent, no matter their number no matter which agent is enabled, which reduces the operational complexity to $\mathcal{O}_c=1$. This approach brings the MAP placement problem to a new dimension where the issue is no longer to determine the architecture of the models but the processing of observations and cooperation. Then, contrary to \textbf{\texttt{C-MADRL}}, where each agent can be distinguished fundamentally by its model, this approach introduces the new challenge of distinguishing agents based solely on observations.
To achieve this, during the training phase \cite{Catte2023}, the federated mechanism retrieves the weights of all the agent models $w_i(t)$ to average them and updates the agent models with a proportion rate $\alpha_f$ every $\tau_f$: $w(t) = \alpha_f \times w(t) + \frac{(1-\alpha_f)}{M_s(t)}\times\sum_i w_i(t)$.
In the proposed solution, parameters $\alpha_f$ and $\tau_f$ ensures the model stability while generalizing, avoiding lack of convergence.% and slowing down. 

The federation of models during the training guarantee that the resulting policy is transferable irrespective of the scenario and the number of deployed MAPs. 

To assess the performance of the aforementioned approaches, we introduce a new metric, termed the operational efficiency, which we define as $\eta(t)=\frac{R(t)}{\mathcal{O}_c}$, where $\mathcal{O}_c$ is the operational complexity defined by the number of different policies to maintain for achieving $R(t)$ (see Table \ref{tab:1}).
%==============================================================================================%
%                                   SIMULATION                                                 %
%==============================================================================================%
\section{Numerical Results}
\begin{table}[!t]
\caption{Simulation Parameters}
    \centering
    \label{table12}
    \scalebox{\tablescale}{
    \begin{tabular}{l||c|c }
    \hline
    \textbf{Channel Parameters} & \textbf{IAB donor} & \textbf{MAP} \\
      \hline
      Carrier Frequency $f_{c}$ & $2\GHz$ & $28\GHz$\\
      Antenna Aperture Angle & $180°$ & $90°$ \\
      Shadowing Variance $\sigma_l^2$ & $3\dB$ & $12\dB$\\
      Antenna Gain & $17~\dBi$ & Directive \cite{Catte2023} \\
      Beam Forming & $K_0=\infty$ & $K_i=10$ \\
      Thermal Noise $N_{0}$ & \multicolumn{2}{c}{$-174\dBm/\Hz$} \\
      Small-Scale fading ($m$-Nakagami) &\multicolumn{2}{c}{$m=3$}\\
      $d_0$ & \multicolumn{2}{c}{$10$}\\
      Bandwidth partition & \multicolumn{2}{c}{$0.75$}\\
      $\Delta \ell$ & \multicolumn{2}{c}{$5$}\\
      $\{\tau_f,\alpha_f\}$ & \multicolumn{2}{c}{$\{5000,0.5\}$}\\
      System Bandwidth $B$ & \multicolumn{2}{c}{$500\MHz$} \\
      Learning rate & \multicolumn{2}{c}{$10^{-4}$}\\
      $\gamma$ & \multicolumn{2}{c}{$0.6$} \\
      $\{N_{i, {\rm UE}}, N_{i,{\rm MAP}}\}$ & \multicolumn{2}{c}{$\{15,5\}$}\\
      $D_i(t)$ ($k$-Poisson distribution) & \multicolumn{2}{c}{$k=1\Gbps$}\\
      \hline
    \end{tabular}}
    \label{tab:channel}
\end{table}

In this section, we evaluate the performance of our two-level hierarchical framework in a dynamic 5G network. To do so, following our previous work \cite{Catte2023}, we train policies using actor-critic framework with proximal policy optimization (PPO) \cite{sana2021transferable}. We refer readers to \cite{Catte2023} for detailed description. %, where we set PPO-clips $(\epsilon_1, \epsilon_2) = (0.01,0.5)$.

For the \textbf{\texttt{CODEBOOK}} construction, we train a set of model for scenarios with $\{2,3,4\}$ MAPs. This approach may be assimilated to the standard \textit{state-of-the-art} approach with specialized models that do not take into account a variable number of MAP.
For the exploitation phase, when $M_s(t)>4$, a random model is sampled from the $M_s(t)=4$ codebook. For the \textbf{\texttt{C-MADRL}} and the \textbf{\texttt{F-MADRL}} training, we randomly deploy from $2$ to $5$ MAPs on a random locations sampled in a $200\m$ by $200\m$ area, where $K(t)=25$ UEs are deployed. Table \ref{table12} summarizes simulation parameters. 

\begin{figure}[!t]
\centering
\includegraphics[width=0.85\columnwidth]{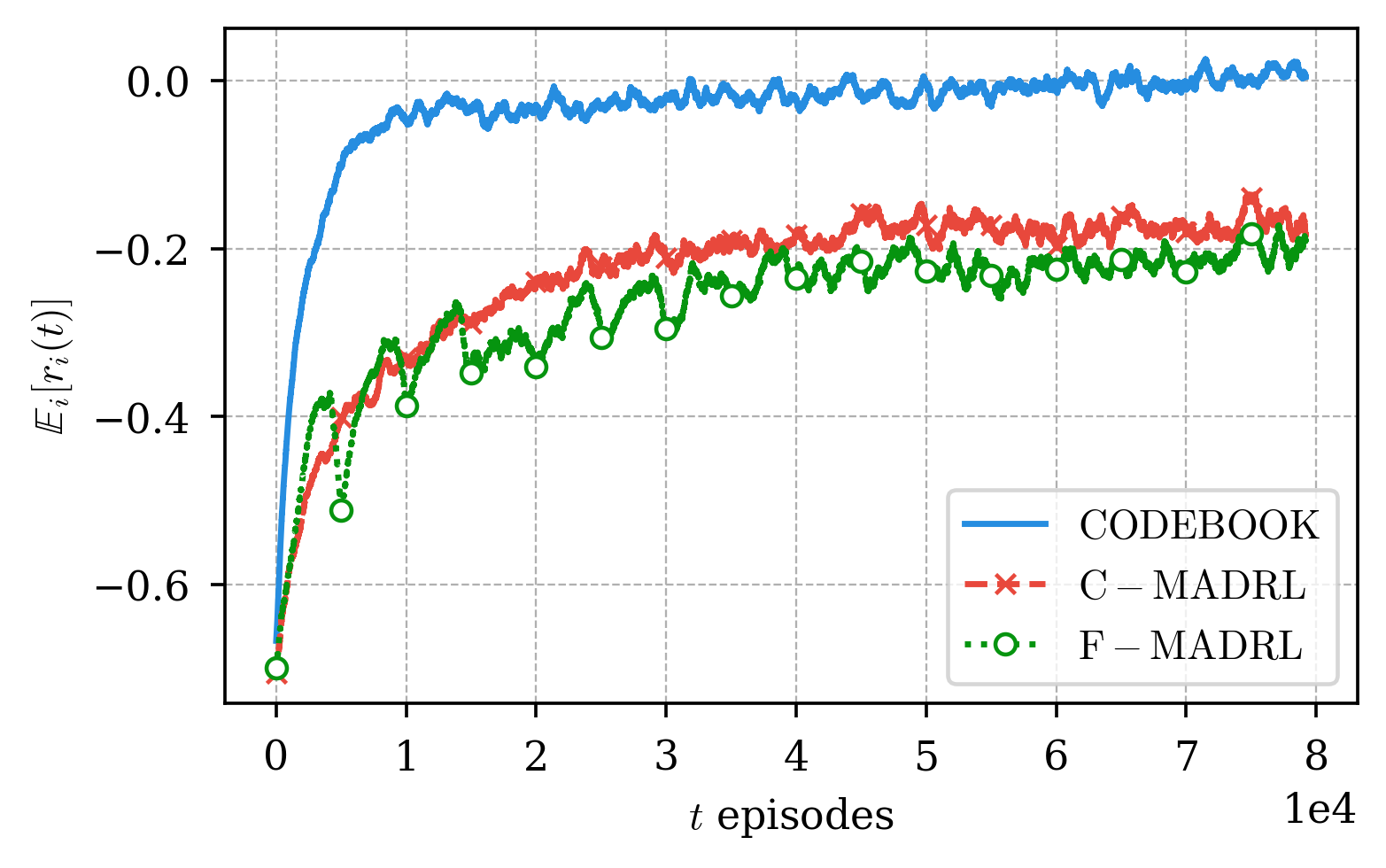}
\caption{Hierarchical Learning convergence of low-level policies.}
\label{fig:reward}
%\vspace{-0.3cm}
\end{figure}

\vspace{0.2cm}
\noindent
\textbf{Federation Policy Convergence.}
To begin, we assess convergence performances of proposed benchmarks. Fig. \ref{fig:reward} shows the rolling averaged reward over a $500$-sized window and over all agents. Under the constraints of a single policy, the \textbf{\texttt{F-MADRL}} solution is able to acquire the capacity to cooperate within a single policy as it have the same convergence than the \textbf{\texttt{C-MADRL}} and \textbf{\texttt{CODEBOOK}} approaches. Though there are drops in reward due to the federation mechanism, it stabilizes during training, confirming the acquisition of cooperation capacity in one single policy. However, due to the generalization capability provided by the federation, the observed reward is lower compared to the specialized \textbf{\texttt{CODEBOOK}} approach, which is specialized for a every scenario.

\begin{figure}[!t]
\includegraphics[width=\columnwidth]{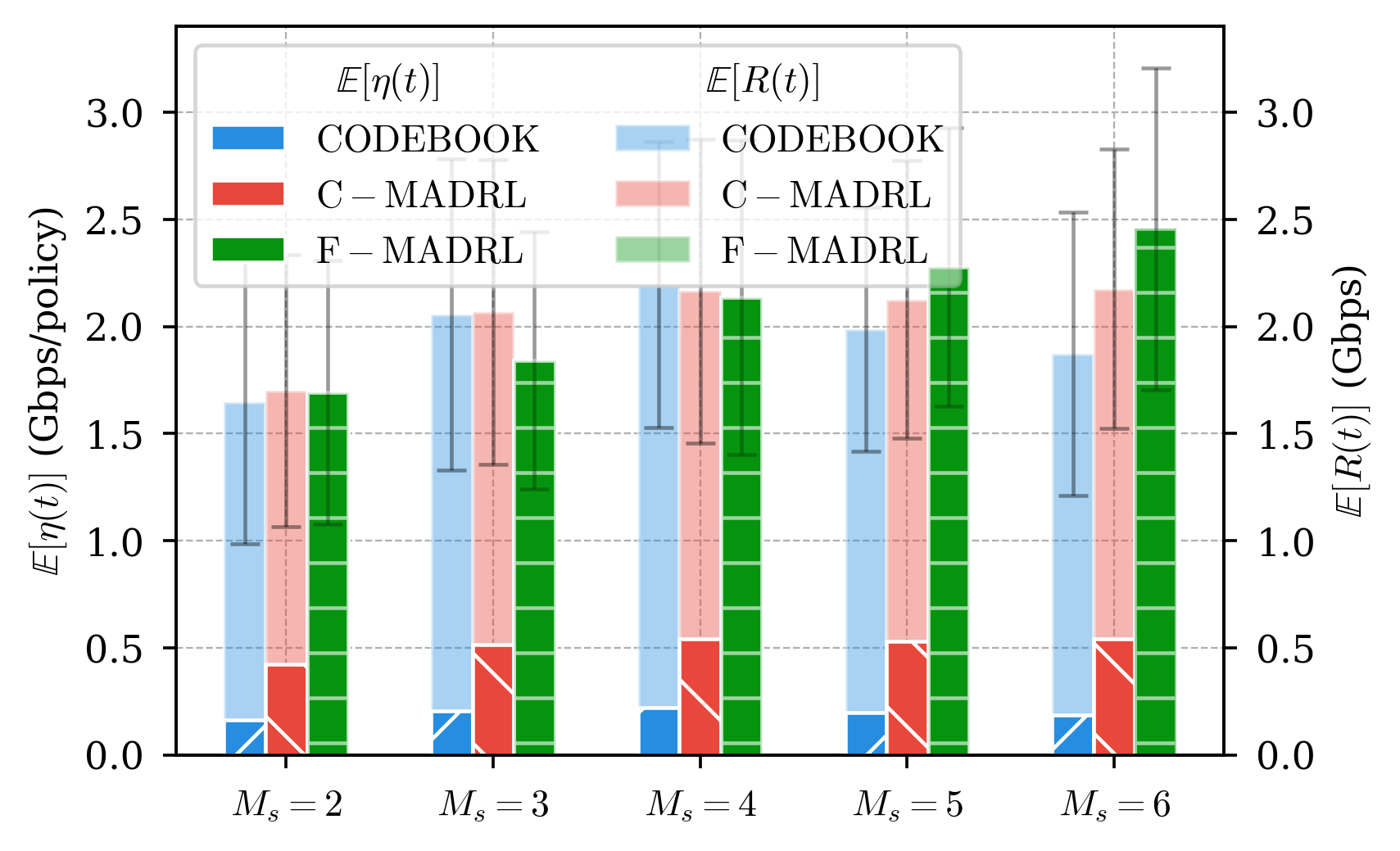}
\caption{Performance comparison between proposed approaches.}
\label{fig:bar11}
\end{figure}

\vspace{0.2cm}
\noindent
\textbf{Federation for Generalization.}
We examine every MADRL generalization ability in the dynamic 5G network. For $200$ configurations that last $T_l=100$ iterations, we deploy now $K(t)=60$ UEs, which does not correspond to any training scenario and $M_s = K(t)/K_i$ MAPs at $t=0$. UEs now follow a random way-point centroid mobility at $0.8\mathrm{m/s}$ with a blockage probability of $0.5$ that leads to a variable total number of connected UE and MAPs between each episodes. As every model has not been trained with specific mobility model, it is able to support multiple type of mobility.
Fig. \ref{fig:bar11} compares the averaged sum-rate achieved $\mathbb{E}[R(t)]$ for different network scenarios. Here, the \textbf{\texttt{CODEBOOK}} approach suffers from a drop of performance in unseen scenarios, while the \textbf{\texttt{F-MADRL}} continuously increases and scales with the network with a $31\%$ improvement with $M_s(t)=6$, while the \textbf{\texttt{C-MADRL}} stabilizes its performances with $M_s(t)\geq 4$.
Most importantly, we examine the operational efficiency $\mathbb{E}[\eta(t)]$ and we observe that the loss of performance in the training process to increase the generalization capacity of the \textbf{\texttt{F-MADRL}} single policy is largely compensated by its cost of exploitation.

\vspace{0.2cm}
\noindent

\begin{figure}[!t]
\vspace{-0.2cm}
\includegraphics[width=\columnwidth]{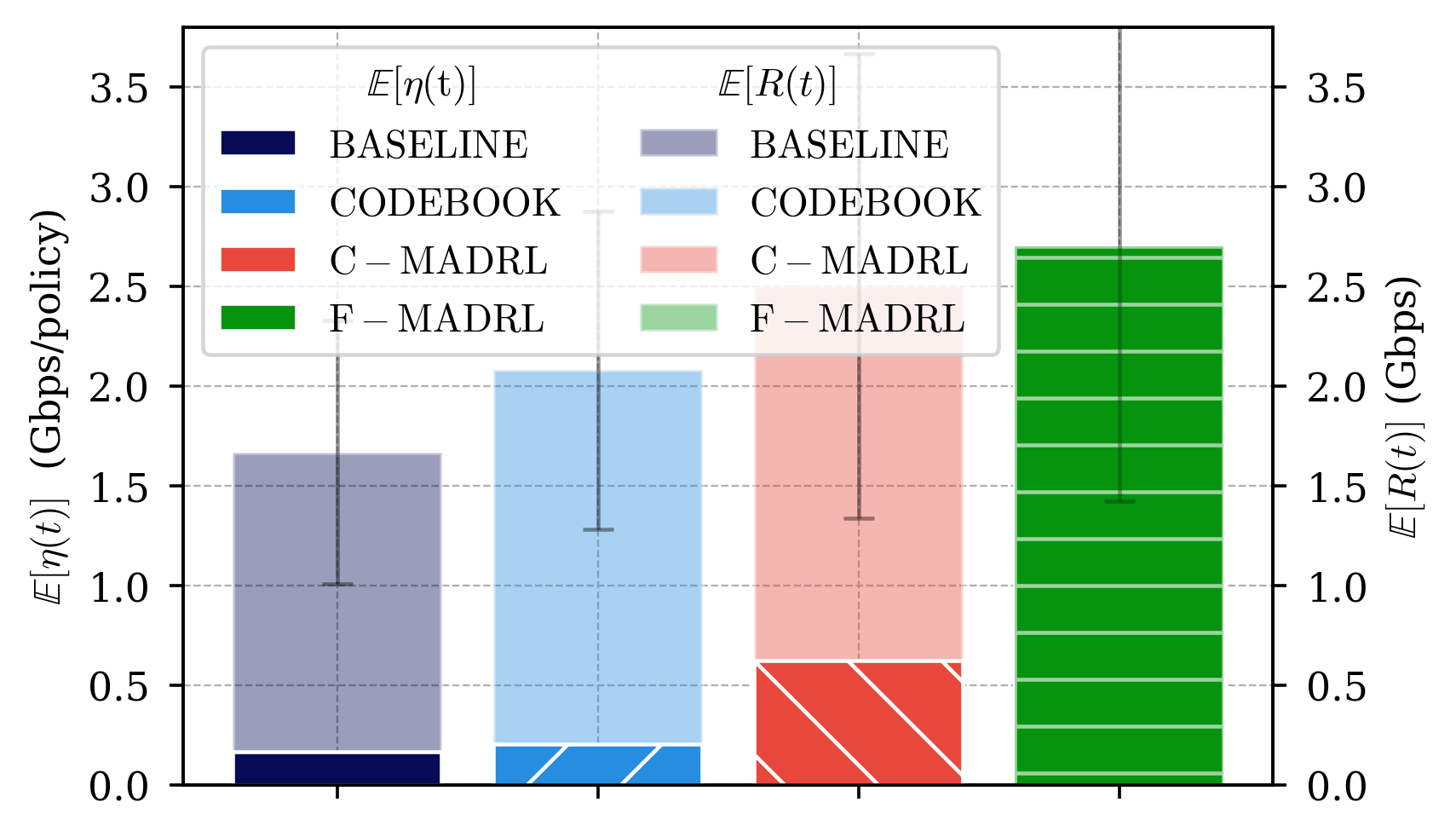}
\caption{Performance comparison compared to the baseline.}
\label{fig:bar21}
\vspace{-0.5cm}
\end{figure}

\vspace{0.2cm}
\textbf{Performance Comparison with Fixed Number of MAPs.}
Here, we use the proposed \textbf{\texttt{trade-off algorithm}} to dynamically manage the MAP number within the same episode every $t_n = 10$. We set inertia thresholds to $\varPhi_{i,\max}=6\times 10^3$, $\varPhi_{i,\min}=0$ and MAP $i$ loads thresholds $K_{i,\min}=2$, $K_i(t)=10$. For each threshold, the trade off of each MAP $\theta_i(t)$ is increased or decreased by 1.
Fig. \ref{fig:bar21} compares $\mathbb{E}_i[R(t)]$ and $\mathbb{E}_i[\eta(t)]$ for all network configurations encountered and MAP $i$.
Thus, compared to the state-of-art \textbf{\texttt{BASELINE}}, which consider a codebook with a \textbf{fixed} number of MAP deployed within an episode, our two-level hierarchical framework achieves an increase of $62\%$ of the averaged sum-rate while demonstrating its operational efficiency. Moreover, the introduction of a dynamic number of MAP for the \textbf{\texttt{CODEBOOK}} approach results in a $24\%$ increase of $\mathbb{E}[R(t)]$, which confirms the need for MAP number adjustment in Multi-MAP networks to meet 5G ambitions. This sum-rate increase can be explained by the better management of UE mobility and interference. As a result, our solution is able to guarantee a better performance even with a high number and density of UEs.

%==============================================================================================%
%                                         CONCLUSION                                          %
%==============================================================================================%
\section{Conclusion}
This study proposes a scalable and distributed solution for determining the optimal placement and number of MAPs in a dynamic 5G network with IAB constraint. The solution utilizes a two-layer hierarchical approach where MAPs decide on their number and optimize backhaul connectivity while autonomously reconfiguring the network. Numerical evaluations show up to $62\%$ network sum-rate increase and improved operation efficiency compared to a state-of-the-art baseline. The proposed solution removes the constraint for a fixed number of deployed MAPs, paving the way for more realistic multi-agent systems with a varying number of agents.

\section*{Acknowledgment}
This work was supported by the European Union H2020 Project \mbox{DEDICAT 6G} under grant no. 101016499. The contents of this publication are the sole responsibility of the authors and do not in any way reflect the views of the EU.

\bibliographystyle{ieeetr}
\bibliography{biblio}

\end{document}